\documentclass[twocolumn,10pt,amsmath,amssymb,prc,aps,superscriptaddress]{revtex4-2}
\usepackage{graphicx}
\graphicspath{{figs/}}
\usepackage{bm}
\usepackage[T1]{fontenc}
\usepackage{hyperref}
\usepackage[normalem]{ulem}
\hypersetup{
 colorlinks  = true, 
 urlcolor   = black, 
 linkcolor  = blue, 
 citecolor  = red  
} 
\usepackage[dvipsnames]{xcolor} 
\usepackage{tikz}
\usepackage{pgfplots}

\newcommand{\UJ}{\mbox{Institute of Theoretical Physics, Jagiellonian University, ul. prof. S. Łojasiewicza 11 PL-30-348 Kraków, Poland}}

\begin{document} 
\title{Engineering interactions shape in resonantly driven bosonic gas}

\author{Damian W{\l}odzy\'nski} \email{damian.wlodzynski@uj.edu.pl} 
\affiliation{\UJ}
\author{Krzysztof Sacha} \email{krzysztof.sacha@uj.edu.pl}
\affiliation{\UJ} 

\date{\today}
\begin{abstract}
	
In systems with fast periodic driving, there are special subsets of (resonant) states, which behavior can be described with effective, time-independent Hamiltonian in a rotating reference frame. Here, we show that experimentally feasible system of ultracold bosonic atoms on a ring with rapidly oscillating scattering length can be used to simulate time-independent two-component atomic mixture with exotic, long-range interactions.
\end{abstract}
\maketitle

The idea behind quantum simulators is to use systems with high degree of experimental control to simulate less-accessible quantum systems. Good candidates for such simulators are systems of trapped ions\cite{Blatt2012}, Rydberg atoms\cite{Weimer2010}, superconducting circuits\cite{Houck2012}, as well as ultracold atoms in optical lattices\cite{science.aal3837}. However even without the optical lattice, ultracold gases are a powerful tool to simulate various quantum systems \cite{Altmeyer2007, Ray2014, Meinert2017, Kwon2020} due to precise control over shape of the trapping potential and interaction strength\cite{pethick2008bose}. In the context of this Letter, it is important, that one can create effectively one-dimensional systems by applying tight-enough confinements in perpendicular directions\cite{Gorlitz2001}. Special case of 1D trap is a ring trap, which put atoms in uniform potential with periodic boundary conditions. Experimentally, such trap is created by combining two trapping potentials: an optical trap confining the system to two dimensions and RF-dressed magnetic trap \cite{Heathcote_2008,Herve_2021}. On the other hand the ultracold atoms typically interact though contact interactions, which depends on the single parameter, s-wave scattering length. The value of the scattering length can be control either with magnetic field using Feshbach resonance or, in the case of quasi-1D systems, by changing trapping depth in perpendicular directions\cite{PhysRevLett.81.938}.

An even wider range of physical systems to simulate can be reached by utilizing the time dimension. For ultra-cold atoms in optical lattice, Floquet enginering can be used to modify solid-state properties of the system \cite{Eckardt2005,oka2019floquet,weitenberg2021tailoring}. This technique has been extensively used in the experiments, mostly with temporal modulation of the optical lattice \cite{Lignier2007,Jotzu2014,Gorg2019,Wintersperger2020}, but not only \cite{Meinert2016}. If atoms are not placed in an optical lattice potential but are instead confined in a trapping potential, solid-state-like behavior can still emerge when a time-periodic perturbation is applied that is resonant with the motion of the trapped atoms \cite{Guo2013,Sacha15_SciRep}. In the theoretical description of such resonant driving, secular terms appear that act as effective external potentials for the atoms. These potentials can be shaped almost arbitrarily, enabling the formation of phase space crystals or time-crystalline structures \cite{SachaTC2020,GuoBook2021}.

Here we consider system with periodically changing strength of contact interactions, which has been of interest in various other theoretical works\cite{Ramos2008,Pollack2010,Rapp2012,Wang2014,Wang2020,Xie_2024}. When the atomic scattering length is modulated at a frequency resonant with the atoms’ motion, it becomes possible to control the effective interaction between two atoms. This opens the door to realizing Anderson molecules \cite{PhysRevA.103.023320} or topological molecules—bound states \cite{PhysRevResearch.6.043173} that arise, respectively, from effective interactions that vary randomly with interatomic distance, and from topologically protected edge states. A recent experimental result for such driving\cite{guthmann2025} show that the system presented in this work is feasible for future experiments.

In this Letter we show how a resonantly driven single-component system of trapped bosons, which can be created in an experiment, can be used to simulate time-independent two-component mixture with exotic intercomponent interactions which otherwise would be impossible to achieve in a physical system.

We consider a system of $2N$ ultracold bosonic atoms on a ring. The system is quasi-1D due to strong confining potential in perpendicular directions. For convenience, we choose units such that $\hbar=1$, the mass of atoms $m=1$ and the size of the ring $L=2\pi$. Interatomic contact interactions are modeled with Dirac delta distribution. We assume that the strength of the interactions (i.e. the s-wave scattering length of atoms) is periodically modulated in time with the period $T=2\pi/\omega$. In the second quantization, the Hamiltonian has a form:
\begin{equation}
	\label{init_ham}
\begin{aligned}
\hat{\cal H}(t) &=\! \int\displaylimits_0^{2\pi}\!dx\ \hat{\Psi}^\dagger(x)\!\left(\!-\frac{1}{2} \frac{d^2}{dx^2} \right)\!\hat{\Psi}(x)+\\
&+ \pi g(\omega t)\! \int\displaylimits_0^{2\pi}\!dx\ \hat{\Psi}^\dagger(x) \hat{\Psi}^\dagger(x) \hat{\Psi}(x) \hat{\Psi}(x)
\end{aligned}
\end{equation}
where $\hat{\Psi}(x)$ and $\hat{\Psi}^\dagger(x)$ are bosonic field operators. The strength of the interactions $2\pi g(\omega t)$, proportional to the s-wave scattering length, can be described as a Fourier series:
\begin{align}
	\label{g_of_t}
	g(\omega t) = \sum_{\substack{m=-K \\ m \neq 0}}^{K}  g_m e^{im \omega t}
\end{align}
where we assume that there is a high frequency cutoff $K$ and that $g_0=0$ (we will discuss $g_0 \neq 0$ case later on).

\begin{figure}
	\centering
	\includegraphics[width=0.3\textwidth]{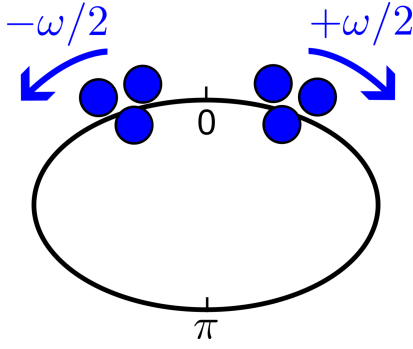}
	\caption{Visualization of the analyzed system of identical bosons. Half of the total number of atoms move clockwise, and half moves counterclockwise along the ring with a frequency $\omega/2$. The s-wave scattering length of atoms is periodically modulated with a frequency $\omega$. Although the original interactions between atoms are contact interactions, the system allows the simulation of a two-component boson gas with arbitrary long-range inter-component interactions.}
	\label{Fig1}
\end{figure}

Since the Hamiltonian is periodic in time, any solution can be decomposed into the Floquet states which evolve periodically in time apart from phase factors. Here we are looking for specific subspace of solutions, which are in the $1:1$ resonance with the driving and have stationary center of mass (i.e. the total momentum of the system is zero). Classically that would correspond to the state with half of the atoms moving clockwise on the ring with angular velocity $\omega/2$, while the other half move analogically counterclockwise (Fig. \ref{Fig1}). If the atoms were distinguishable, one could find these states by transforming every atom to the appropriate rotating frame and looking for solutions with low energy of an effective Hamiltonian \cite{PhysRevResearch.6.043173}. However, since these are identical bosons, that would not preserve their exchange symmetry. One can avoid this issue by transforming every particle to "both" rotating frames.

The idea behind the transformation is to make it momentum dependent. If atom has a positive momentum, it is transformed to the frame rotating clockwise, otherwise to the counterclockwise. For the mathematical description, we first move to the momentum space:
\begin{align}
	\hat{\cal H} = \sum_{k=1}^\infty \frac{k^2}{2} \hat{a}^\dagger_k \hat{a}_k+\frac{g(\omega t)}{2}  \sum_{klmn} \hat{a}^\dagger_n \hat{a}^\dagger_m \hat{a}_k \hat{a}_l \delta_{n+m-k-l} 
\end{align}
and then perform the transformation
\begin{align}
	\hat{\cal H}_{\text{rot}}&= \hat{U}\hat{\cal H}\hat{U}^{-1} + i (\partial_t\hat{U}) \hat{U}^{-1}\\
	\hat{U}&=e^{i\frac{\omega}{2} t \sum_k |k| \hat{a}_k^\dagger \hat{a}_k}
\end{align}
As a result we obtained:
\begin{align}
	\begin{aligned}
	\hat{\cal H}_\text{rot}&=  \sum_k \frac{1}{2}(|k|-\frac{\omega}{2})^2 \hat{a}_k^\dagger\hat{a}_k+\\
	&+\frac{g(\omega t)}{2} \sum_{klmn} \hat{a}^\dagger_n \hat{a}^\dagger_m \hat{a}_k \hat{a}_l   e^{i\frac{\omega}{2} t(|n|+|m|-|k|-|l|)}\delta_{n+m-k-l}
	\end{aligned}
\end{align}
with additional constant $-N\omega^2/4$ which was excluded. The resulting Hamiltonian is non-local in the position space, but it turns out to not be an issue in the further analysis.

According to the Floquet theorem, the evolution operator for an integer number of periods can be described as $\hat{U}(nT+t_0,t_0)=e^{i\hat{H}_F nT}$, where time independent operator $\hat{H}_F$ is called the Floquet Hamiltonian. Therefore, performing stroboscopic measurements on the original system (\ref{init_ham}) is mathematically equivalent to probing time-independent system described by $\hat{H}_F$. The Floquet Hamiltonian for the analyzed system cannot be obtained analytically, but can be written as a series in powers of $1/\omega$, called the Magnus expansion \cite{Blanes_2010}. In this letter we are interested in the case of very large driving frequency $\omega \gg  gN, K, E_{\text{sys}}$ in the used dimensionless units, where $E_{\text{sys}}$ is the average energy of the system in the rotating frame and $g=\max|g_m|$. Under this assumptions, we can simplify the Floquet Hamiltonian to the first term of the expansion, which is known as a secular approximation. As a result we get an effective Hamiltonian:

\begin{align}
	\label{ham_eff}
	\begin{aligned}
	\hat{\cal H}_\text{eff} &= \frac{1}{T} \int_0^{T}\!\!dt\ \hat{\cal H}_\text{rot} =  \sum_k \frac{1}{2}(|k|-\frac{\omega}{2})^2 \hat{a}_k^\dagger\hat{a}_k\\
	&+ \sum_{\substack{q=-K \\ q \neq 0}}^{K} \frac{g_q}{2} \sum_{klmn} \hat{a}^\dagger_{n}\hat{a}^\dagger_{m}\hat{a}_{k}\hat{a}_{l}\delta_{n+m-k-l}\delta_{|n|+|m|-|k|-|l|+2q}
	\end{aligned}
\end{align}

\begin{figure}
	\centering
	\includegraphics[width=0.4\textwidth]{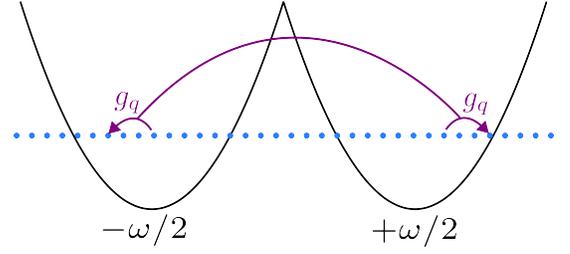}
	\caption{Visualization of the different terms in the Hamiltonian (\ref{double_well}). Blues dots illustrate momentum states (chain in the momentum space), while solid black line show single particle kinetic energy (double well in the momentum space). Couplings between states is indicated with purple arrows.}
	\label{Fig2}
\end{figure}

The assumption about high frequency cutoff in the driving $K\ll \omega$ allow for even further simplifications. In this case, the subspace of states with all momenta $|k|\approx \omega/2$ is approximately decoupled from the rest of states and can be described by the Hamiltonian:
\begin{align}
	\label{double_well}
	\begin{aligned}
		\hat{\cal H}_\text{eff}  &=  \sum_k \frac{1}{2}(|k|-\frac{\omega}{2})^2 \hat{a}_k^\dagger\hat{a}_k +\\
		&+ \sum_{\substack{q=-K \\ q \neq 0}}^{K} 2g_q \sum_{\substack{k\approx +\omega/2 \\ l\approx -\omega/2}} \hat{a}^\dagger_{l+q}\hat{a}^\dagger_{k-q}\hat{a}_{k}\hat{a}_{l}
	\end{aligned}
\end{align}
However, the eigenstates obtained from this Hamiltonian are guaranteed to be low energy solutions only for sufficiently weak interactions. As a safe assumption we choose $\omega^2 \gg gN^2$, since then kinetic energy from even a single particle far from the minimum cannot be compensated by the interaction term. The effective Hamiltonian (\ref{double_well}) can be visualized as a bosonic chain in a double well potential in the momentum space (Fig. \ref{Fig2}).

Finally, we check how reasonable are our assumptions for experimental setups. In a recent experiment \cite{guthmann2025}, it was possible to perform modulation of the interaction strength with the frequency of the order of MHz. To translate it to our dimensionless units, one has to divide it by $\frac{\hbar}{M R^2}$, where $M$ is atomic mass and $R$ is the radius of the ring. For example, for $R$ of the order of $100 \mu m$ and cesium atoms, we get $\omega \approx 10^5$. Therefore, even for relatively strong interactions $gN=10$, we still get reasonable constraint for number of atoms in the system $N \ll 10^9$. The approximation improves for larger trap and heavier atoms.

An important property of the interactions in Hamiltonian (\ref{double_well}) is that they couple bosons from both the wells in the momentum space (Fig. \ref{Fig2}), but do not depend on initial distance between coupled particles (and therefore also on the distance between the wells). This system is mathematically equivalent to a single well problem with two different atomic species. To show it, we define a new set of creation/annihilation operators:
\begin{align}
	\label{new_op}
	\begin{aligned}
	\hat{a}_{k,\uparrow} &= \hat{a}_{\omega/2+k}\\
	\hat{a}_{k,\downarrow} &= \hat{a}_{-\omega/2+k}
	\end{aligned}
\end{align}
with $|k|\ll \omega/2$, which satisfy bosonic commutation relations $[\hat{a}_{k,\sigma},\hat{a}_{k',\sigma'}^\dagger]=\delta_{kk'} \delta_{\sigma\sigma'}$. With these new operators, the effective Hamiltonian has a form:
\begin{align}
	\begin{aligned}
	\hat{\cal H}_\text{eff} &= \sum_{k,\sigma=\uparrow,\downarrow} \frac{k^2}{2} \hat{a}_{k,\sigma}^\dagger \hat{a}_{k,\sigma} +\\
	&+ \sum_{\substack{q=-K \\ q \neq 0}}^{K} 2g_q \sum_{k,l} \hat{a}^\dagger_{l+q,\downarrow}\hat{a}^\dagger_{k-q,\uparrow}\hat{a}_{k,\uparrow}\hat{a}_{l,\downarrow}
	\end{aligned}
\end{align}
Next, we define bosonic field operators corresponding to the operators(\ref{new_op}):
\begin{align}
	\hat{\Psi}_\sigma(x) := \sum_k \frac{1}{\sqrt{2\pi}} e^{ikx} \hat{a}_{k,\sigma}
	\label{field_op2}
\end{align}
It should be emphasized that by themselves they do not have physical interpretation of the annihilation operator in the position space of our original problem, but they are related to it by equation: $\hat{\Psi}(x)=e^{i\omega x/2}\hat{\Psi}_\uparrow(x)+e^{-i\omega x/2}\hat{\Psi}_\downarrow(x)$. However in this description it is ease to go back to the lab frame, because $\hat{U}\hat{\Psi}_{\uparrow}(x) \hat{U}^{-1} = \hat{\Psi}_{\uparrow}(x-\omega t/2)$ and $\hat{U}\hat{\Psi}_{\downarrow}(x) \hat{U}^{-1} = \hat{\Psi}_{\downarrow}(x+\omega t/2)$. 

With the help of (\ref{field_op2}), the effective Hamiltonian takes the form:
\begin{align}
	\label{ham_eff_pos}
	\begin{aligned}
	\hat{\cal H}_\text{eff} &= \sum_{\sigma=\uparrow,\downarrow} \int_0^{2\pi}\!\!\!\!dx \hat{\Psi}_\sigma^\dagger(x) \left(-\frac{1}{2}\frac{\partial^2}{\partial x^2}\right)\hat{\Psi}_\sigma(x)+\\
	&+2\int_0^{2\pi}\!\!\!\!dx\int_0^{2\pi}\!\!\!\!dy \hat{\Psi}_\uparrow^\dagger (x) \hat{\Psi}_\downarrow^\dagger(y) g(x-y) \hat{\Psi}_\downarrow(y) \hat{\Psi}_\uparrow (x).
	\end{aligned}
\end{align}
This Hamiltonian describes a two-component bosonic system on a ring in the absence of intra-component interactions but in the presence of the inter-component interactions described by the function $g(x-y)$. In the original Hamiltonian (\ref{init_ham}), this function describes modulation of the interaction strength in time. Here however it is responsible for the shape of the interaction potential. Therefore by choosing proper driving function $g(t)$ in the original problem, we control shape of the interaction potential in this effective description. That allows for the simulation of exotic long-range interactions that would otherwise be impossible to achieve. In the experiment, the information about the effective system would be obtained through stroboscopic measurements in the original system. The applied measurement technique should be able to distinguish between particles with opposite momenta.

So far we were assuming that there is no time-independent component $g_0$ in the strength of the original contact interactions in (\ref{g_of_t}). If this component is included, it does not affect the inter-component interactions in the effective Hamiltonian, because: $\int_0^{2\pi}\!\!\!\!dx\int_0^{2\pi}\!\!\!\!dy \hat{\Psi}_\uparrow^\dagger (x) \hat{\Psi}_\downarrow^\dagger(y) g_0 \hat{\Psi}_\downarrow(y) \hat{\Psi}_\uparrow (x) = g_0 N^2$, which is just a constant term. However, it results in additional terms when going from Eq.(\ref{ham_eff}) to Eq.(\ref{double_well}), which describe interactions between atoms in the same wells in the momentum space. Therefore there is an additional term in the effective Hamiltonian (\ref{ham_eff_pos}): $+\pi g_0\sum_{\sigma=\uparrow,\downarrow}\! \int\displaylimits_0^{2\pi}\!dx\ \hat{\Psi}_\sigma^\dagger(x) \hat{\Psi}_\sigma^\dagger(x) \hat{\Psi}_\sigma(x) \hat{\Psi}_\sigma(x)$. As a result, we now can control not only the shape of the inter-component interaction potential but also the strength of the contact intra-component interactions, making our system even more general tool for realization of systems with interactions that even do not exist in nature.

We should also discuss the case when the original interactions between atoms are not perfectly described as contact interactions, which can happen if the atoms are moving with very large velocities along the ring. For finite-range interactions, which do not change the internal states of the atoms, the presented method can still be utilized. If the interaction in momentum space has a form: $g(\omega t)\sum_q f_q \sum_{k,l} \hat{a}^\dagger_{l+q}\hat{a}^\dagger_{k-q}\hat{a}_{k}\hat{a}_{l}$ and each term $f_q$ between $q=-K$ and $q=K$ is non-zero, then every shape of the inter-component interaction potential in the effective description (\ref{ham_eff_pos}) can be realized by appropriately choosing $g_m$ coefficients in (\ref{g_of_t}).

Finally, we would like to emphasize that the interaction potential in (\ref{ham_eff_pos}) can be modified during the course of the experiment. It is sufficient to change, at a chosen moment, the shape of the function $g(\omega t)$ describing the modulation of the s-wave scattering length, and the interaction potential $g(x-y)$ will change accordingly.

One example of an exotic strongly correlated system which could be probed using this approach is many-body generalization of so-called Anderson complexes \cite{PhysRevA.103.023320}. The system would be simulated using periodically repeated disordered driving $g(\omega t)\propto\sum_m e^{im \omega t+i\phi_m}$ ($\phi_m$ -- random phases), which would translate to the disordered interactions in the effective model. It has been shown in 2-body case, that such interaction leads to Anderson-like localization in the relative distance between atoms. Nevertheless, it remains an open question how these disordered interactions would affect the many-body system, especially in the strong interactions regime. The presented approach would allow such system to be probed experimentally.

In summary, we have presented a concept that enables the simulation of a two-component Bose gas of ultracold atoms with arbitrary long-range inter-component interactions using a single-component atomic gas moving on a ring. A key element of this approach is the periodic modulation of the atoms’ s-wave scattering length, resonant with their motion along the ring. This technique becomes a universal tool for the experimental realization of exotic and potentially strongly correlated systems.

\bibliographystyle{apsrev4-1}
\bibliography{bib}

\end{document}